\documentclass[conference]{IEEEtran}

\usepackage{amssymb}
\usepackage{amsmath}
\usepackage{amsfonts}
\usepackage{dsfont} 
\usepackage[utf8]{inputenc}
\usepackage{amsthm}
\usepackage{cite}
\usepackage{subfigure}
\usepackage{color}
\usepackage{times}
\usepackage{graphicx}
\usepackage{psfrag}

\newcommand{\bm}{\mathbf}
\newcommand{\be}{\begin{equation}}
\newcommand{\ee}{\end{equation}}
\newcommand{\bea}{\begin{eqnarray}}
\newcommand{\eea}{\end{eqnarray}}

\newcommand{\p}{{\bm p}}

\newcommand{\ba}{{\bm a}}

\newcommand{\bA}{{\bm A}}
\newcommand{\bI}{{\bm I}}

\newcommand{\bF}{{\bf F}}

\newcommand{\bD}{{\bf D}}
\newcommand{\bC}{{\bf C}}

\newcommand{\bP}{{\bf P}}

\newcommand{\bh}{{\bf h}}
\newcommand{\bz}{{\bf z}}

\newcommand{\BDFT}{{\boldsymbol{\mathcal F}}}

\IEEEoverridecommandlockouts
\begin{document}

\title{Spectrally Efficient Pilot Structure and Channel Estimation for Multiuser FBMC Systems}

\author{\normalsize Hamed Hosseiny$^\dagger$, Arman Farhang$^*$ and Behrouz Farhang-Boroujeny$^\dagger$  
\\$^\dagger$ECE Department, University of Utah, USA, \\
$^*$Department of Electronic Engineering, Maynooth University, Ireland. \\
Email: \{hamed.hosseiny, farhang\}@utah.edu, \{arman.farhang\}@mu.ie}
\maketitle

\begin{abstract}
In this paper, we consider channel estimation problem in the uplink of filter bank multicarrier (FBMC) systems. We propose a pilot structure and a joint multiuser channel estimation method for FBMC. Opposed to the available solutions in the literature, our proposed technique does not rely on the flat-channel condition over each subcarrier band or any requirement for placing guard symbols between different users' pilots. Our proposed pilot structure reduces the training overhead by interleaving the users' pilots in time and frequency. Thus, we can accommodate a larger number of training signals within the same bandwidth and improve the spectral efficiency. Furthermore, this pilot structure inherently leads to a reduced peak-to-average power ratio (PAPR) compared with the solutions that use all the subcarriers for training. We analytically derive the Cram\'er-Rao lower bound (CRLB) and mean square error (MSE) expressions for our proposed method. We show that these expressions are the same. This confirms the optimality of our proposed method, which is numerically evaluated through simulations. Relying on its improved spectral efficiency, our proposed method can serve a large number of users and relax pilot contamination problem in FBMC-based massive MIMO systems. This is corroborated through simulations in terms of sum-rate performance for both single cell and multicell scenarios.
\end{abstract}

\section{Introduction}
Filter Bank Multi-Carrier (FBMC) has been considered as a promising candidate waveform for the future wireless systems \cite{farhang2011ofdm,nissel2017filter}. As a multiuser system, massive MIMO is becoming a reality, and the application of FBMC to massive MIMO channels has been recently studied in \cite{farhang2014filter}. The authors in \cite{farhang2014filter} showed that self-equalization/channel-flattening effect makes FBMC an interesting choice for signaling over massive MIMO channels. However, massive MIMO highly relies on accurate channel estimates to deliver all of its promising benefits. Furthermore, many of the emerging applications in future wireless networks require ultra-reliable low-latency communications (URLLC), \cite{schulz2017latency}. This necessitates the need for highly accurate channel estimation techniques with minimal training overheads especially in multiuser scenarios. Hence, efficient multiuser channel estimation is the topic of interest to this paper.

Orthogonality in FBMC only holds in the real field. This makes channel estimation in FBMC more complex than in orthogonal frequency division multiplexing (OFDM). The existing channel estimation methods for FBMC in the literature are mainly based on the interference approximation method (IAM), which was first introduced in \cite{lele2008channel} and only holds when the channel response over each subcarrier band can be approximated by a flat gain. IAM is a frequency domain channel estimation technique and requires maximum channel delay spread to be much smaller than the symbol interval. Thus, when this condition does not hold, IAM leads to inaccurate channel estimates. To avoid this issue, time domain channel estimation techniques were proposed \cite{kong2014time,caus2012transmitter,kofidis2015preamble,singh2019time}. 

The authors in \cite{kong2014time} and \cite{caus2012transmitter} propose time domain channel estimation techniques where guard symbols are required to separate different users' pilots.
This leads to spectral efficiency loss which is not desirable as the number of users increases. An alternative time domain channel estimation method for FBMC and its extension to MIMO channels was proposed in \cite{kofidis2015preamble} and further studied in \cite{singh2019time}. This method considers sending pilots for each user on all the subcarriers where the users' pilots are multiplexed in the code domain to allow sharing the same time-frequency resources for channel estimation. However, this solution suffers from a large amount of computational load at the receiver to demultiplex different users' channel responses based on their code sequences.

In this paper, we propose a pilot structure and a time domain channel estimation method for multiuser FBMC systems. We consider the uplink scenario where each user is transmitting its pilots on a subset of disjoint subcarriers, i.e., different time-frequency resources. Opposed to the existing literature, we propose a pilot structure that interleaves different users' pilots in time and frequency without the need for any guard symbols between them. We also show that for a given user with the channel length $L_u$, utilization of only $L_u$ pilots for channel estimation is sufficient. This allows us to jointly estimate the channel responses of a large number of users at the base station and achieve an improved spectral efficiency. Considering a fixed training power budget per user, this allows us to boost the transmit signal power and achieve more accurate channel estimates in multiuser systems. It is worth to note that, guard symbols are still required to protect preamble from data symbols. Moreover, one may note that reducing the number of pilots in the training signal of each user, leads to a reduced peak-to-average power ratio (PAPR) compared to \cite{kofidis2015preamble} and  \cite{singh2019time} where the pilots are inserted on all the subcarriers. In our proposed channel estimator, we take advantage of the intrinsic interference due to the absence of guard symbols between different users' pilots and jointly estimate all the users' channel impulse responses. We analytically derive the mean square error (MSE) and Cram\'er-Rao lower bound (CRLB) expressions for our proposed technique. Based on our derivations, MSE leads to the same result as CRLB. This confirms the optimality of our proposed channel estimation technique. We assess the efficacy of our proposed method and corroborate the validity of our analytical derivations through numerical simulations. Based on our numerical results, our proposed technique achieves almost the same performance as CRLB. Finally, we show that our proposed solution is very effective for utilization in massive MIMO systems as it provides a great amount of savings in signaling overhead compared to the available solutions and hence can simultaneously serve a large number of users. This leads to a significant relaxation on the pilot contamination problem. To support our claim, we numerically compare the sum-rate performance of our proposed technique against conventional solutions through simulations. Based on this comparison, our solution achieves a substantial amount of improvement in both single cell and multicell massive MIMO setups.

The rest of the paper is organized as follows. Section II presents FBMC principles paving the way towards single user channel estimation in Section III. Section IV presents our proposed pilot structure and multiuser channel estimation technique. Section V discusses the extension of our proposed channel estimation technique to massive MIMO.  Section VI provides numerical analysis of our proposed channel estimation technique while confirming the validity of our claims on spectral efficiency through simulations. Finally, the paper is concluded in Section VII.

\textit{Notations:} Matrices, vectors and scalar quantities are denoted by boldface uppercase, boldface lowercase and normal letters, respectively. $A(m,l)$ represents the element in the $m^{\rm th}$
row and the $l^{\rm th}$ column of $\bA$ and $\bA ^{-1}$ signifies the inverse of $\bA$. $\bI_M$ is the identity matrix of size $M \times M$, and $\bD = {\rm diag} ({\ba})$
is a diagonal matrix whose diagonal elements are formed by the elements of the vector $\ba$. The superscripts $(\cdot)^{\rm T}$, $(\cdot)^{\rm H}$ and $(\cdot)^*$ indicate transpose, conjugate transpose, and conjugate
operations, respectively. The linear convolution is denoted by $\star$. The real and imaginary parts of a complex number are denoted by $\mathfrak{R}\{\cdot\}$ and $\mathfrak{I} \{\cdot\}$, respectively. $\mathds{E}\{\cdot\}$ denotes the expected value of a random variable, and ${\rm tr}\{\cdot\} $ is the matrix trace operator. The notation $\mathcal{CN}(0,\sigma^2)$ represents the
circularly-symmetric complex normal distribution with zero mean and variance $\sigma^2$. Finally, $\delta_{ij}$ represents the Kronecker delta function.

\vspace{-1.5mm}
\section{FBMC Principles}
We consider the baseband equivalent of the staggered multi-tone (SMT) system in discrete time. This modulation scheme divides the bandwidth into $M$ sub-carriers with the normalized bandwidth of $1/M$ each. The real-valued data symbols in SMT are placed on a regular grid in the time-frequency plane with the time-frequency spacing of $T/2$ and $1/T$, respectively. In FBMC systems, overlapping between adjacent time-frequency symbols is allowed. To avoid interference between the adjacent time and frequency symbols, orthogonal/offset quadrature amplitude modulation (OQAM) is deployed. A phase shift of $\pi /2 $ is present between the adjacent symbols, $s_{m,n}$, where $m$ and $n$ are frequency and time indices, respectively. 
In FBMC systems, data symbols in the baseband for a given time slot, $n$, are initially pulse-shaped with a prototype filter $g[k-n\frac{M}{2}]$ and then they are up-converted to different subcarrier frequencies $f_0,\ldots,f_{N-1}$. The length of this filter is considered to be $\kappa M$ where $\kappa$ is the overlapping factor, i.e., the prototype filter spreads over multiple time symbols. Therefore, FBMC transmit signal can be obtained as
\begin{equation}
x[k] = \sum_{m=0}^{M-1} \sum_{n=-\infty}^{ \infty} s_{m,n} g_{m,n}[k] ,
\label{bbeq}
\end{equation}
where $g_{m,n}[k] = g\big[k-n\frac{M}{2}\big] e^{j2\pi mk/M}e^{j\pi (m+n)/2}.$
Pulse-shaping filters $g_{m,n}[k]$, can be thought of as basis functions that are orthogonal in the real field, i.e.,
\begin{equation}
    \mathfrak{R} \bigg\{ \sum_{k=- \infty}^{ \infty} g_{m,n}[k]  g^{*}_{m',n'}[k] \bigg\} = \delta_{mm'} \delta_{nn'}.
\end{equation}

Assuming a time-invariant channel, the received signal at the receiver can be written as
\begin{equation}
    y[k] = h[k] \star x[k] + \eta[k],
    \label{convolutionalform}
\end{equation}
where $h[k]$ represents multi-path channel impulse response and $\eta[k]$ is additive white Gaussian noise (AWGN) with the variance of $\sigma^2$, i.e., $\eta[k] \thicksim \mathcal{CN}(0,\sigma^2)$. 
Using \eqref{bbeq} and \eqref{convolutionalform}, the demodulated signal at symbol $n$ and subcarrier $m$, before taking the real part, can be obtained as $z_{m,n} = \langle y[k], g_{m,n}[k] \rangle$ which can be expanded as

\begin{equation}
\begin{aligned}
&    {z}_{m,n} =         
\sum_{l=0}^{L-1} \sum_{k=-\infty}^{\infty} \sum_{m'=0}^{M-1} \sum_{n' =-\infty}^{\infty} s_{m',n'} g[k-l-n'\frac{M}{2}] \\
 & \times g[k-n\frac{M}{2}] 
   \times e^{j2\pi(m'-m)k/M} e^{j\pi (m'+n'-m-n)/2} 
   \\
   & \times e^{-j2\pi m'l/M}h[l]+\eta_{m,n},
    \end{aligned}
    \label{demod}
\end{equation}
where $\eta_{m,n}= \sum_{k=-\infty}^{\infty} \eta[k] g^*_{m,n}[k]$ is the noise contribution after filtering and phase adjustment. Assuming perfect synchronization and considering presence of an accurate channel estimate at the receiver, the transmit symbols, $s_{m,n}$, can be recovered after performing equalization and real part operations on ${z}_{m,n}$. Since, channel estimation is the main focus of this paper, channel equalization aspects are not discussed. As mentioned earlier, many authors have proposed channel estimation techniques for FBMC systems, \cite{lele2008channel,kong2014time,caus2012transmitter,kofidis2015preamble,singh2019time}. However, these solutions are either limited to frequency flat channels in subcarrier level or they are suffering from a poor spectral efficiency, especially, in multiuser scenarios. Hence, in the following sections, we revisit channel estimation problem in FBMC. Starting from single user scenario, we propose a pilot structure and channel estimation technique that can address the spectral efficiency issues of FBMC in multiuser and massive MIMO systems while maintaining the optimal performance.

\section{Single User Channel Estimation} 
\label{sparsepilot}
In this section, we focus on the time domain channel estimation for single user FBMC systems to pave the way towards our proposed multiuser channel estimation technique in Section~\ref{sec:Multiuser_ch_estimation}. We consider a set of pilots $\p = [p_1,p_2, \ldots ,p_{N_{\rm p}}]^{\textrm{T}}$ and the vector of the demodulated symbols at the receiver as $\bz = [z_{p_1},z_{p_2}, \ldots ,z_{p_{N_{\rm p}}}]^{\textrm{T}}$, where $N_{\rm p}$ is the number of pilots, and $p_i$ and $z_{p_i}$ are the transmit and received symbols at the time-frequency slot $n_{p_i},m_{p_i}$, respectively.  To guarantee complete isolation of training and data symbols, sufficient number of guard symbols are inserted among them. These guard symbols protect training symbols from intrinsic interference issues that arise in FBMC systems \cite{kong2014time}. It is worth noting that the placement and amount of required zero symbols depend on the pulse shaping filter and its overlapping factor. Using \eqref{demod}, we can represent $\bz$ as
\begin{equation}
    \bz = \bA \bh +\mbox{\boldmath$\eta$},
    \label{matform}
\end{equation}
where $\bh=[h[0],\ldots,h[L-1]]^{\rm T}$ represents  the channel impulse response, $\mbox{\boldmath$\eta$}=[\eta_{p_1},\eta_{p_2}, ..., \eta_{p_{N_{\rm p}}}]^{\rm T}$ is the demodulated noise at the time-frequency instance $n_{p_i},m_{p_i}$ and $\bA$ is an $N_{\rm p} \times L$ matrix whose elements $A(m,l)$ can be obtained as
   \begin{equation}
    \begin{aligned}
    A(m,l) &= \!\!\!\! \sum_{k=-\infty}^{\infty} \sum_{m'=0}^{M-1} \sum_{n'=-\infty}^{\infty} \!\! s_{m',n'}  g[k-l-n'\frac{M}{2}]
   g[k-n\frac{M}{2}]  \\
   &\times e^{j2\pi(m'-m)k/M} e^{j\pi (m'+n'-m-n)/2} 
   e^{-j2\pi m' l/M}.
    \end{aligned}
    \label{elementformula}
    \end{equation}
Equation \eqref{matform} is a compact representation of the pilot signals at the output of FBMC receiver before the real-part operation. This formulation decouples the channel and training sequence that is filtered by the transmit and receive filters paving the way towards estimating the wireless channel. 

Hence, the least squares estimate of the channel can be obtained as in \cite{kay1993fundamentals}
    \begin{equation}
       \hat{\bh} = (\bA^{\text{H}}\bA)^{-1} \bA^{\text{H}} \bz  .
       \label{est}
    \end{equation}    

As an alternative to \eqref{est}, channel can be estimated based on minimum mean square error (MMSE) criterion. MMSE surpasses least squares in performance, however, it requires prior knowledge of channel covariance matrix \cite{kay1993fundamentals}.

The proposed time domain channel estimation method in \cite{kong2014time} requires the number of pilots to be equal to the number of subcarriers where sufficient number of guard symbols are inserted to avoid intrinsic interference. This results in bandwidth efficiency loss especially for extension to multiuser channel estimation in the uplink. This is because additional guard symbols are required to separate different users' training signals from one another. Hence, reducing the pilot overhead for channel estimation in FBMC systems is of a paramount importance. To this end, in the following, we show that it is possible to find an accurate channel estimate only by using $L$ rather than $M$ number of pilots. This leads to a significant amount of saving in signalling overhead as the channel length is usually much smaller than the number of subcarriers, e.g. in 3GPP Long Term Evolution (LTE) standard $L\leq 0.1 M$. The additional degree of freedom from reducing the number of pilots enables separation of different users' pilots in multiuser scenarios. \textit{This is one of the main motivations behind our proposed channel estimation method in the following section.}

Considering pilot transmission without intrinsic interference, in the following, we show that the value of the MSE only relates to the total power of pilots as long as the total number of pilots is equal or larger than the channel length. Therefore, we start with the derivation of the mean and covariance of the channel estimates. Using \eqref{est}, we have
 \begin{equation}
\begin{aligned}
\mathds{E}[\hat{\bh}]\! =\! \mathds{E} \big[(\bA^{\text{H}}\bA)^{-1} \bA^{\text{H}} \bz  \big]
\!=\! \mathds{E} \big[(\bA^{\text{H}}\bA)^{-1} \bA^{\text{H}} (\bA \bh+\boldsymbol{\eta})  \big]\! =\!\bh,
\end{aligned}
\end{equation}
 and
\begin{equation}
\begin{aligned}
&{\rm Cov}[\hat{\bh}] = \mathds{E} [(\bh-\hat{\bh}) (\bh-\hat{\bh})^{\text{H}}] 
\\
&= \mathds{E} \big[(\bA^{\text{H}}\bA)^{-1} \bA^{\text{H}} \mbox{\boldmath$\eta$} \mbox{\boldmath$\eta$}^{\text{H}} \bA \big( (\bA^{\text{H}}  
\bA)^{-1} \big)^{\text{H}} \big]= \sigma^2(\bA^{\text{H}}\bA)^{-1}.
\end{aligned}
\end{equation}
This implies that
\begin{equation}
\begin{aligned}
\hat{\bh} \thicksim \mathcal{CN}(\bh, \sigma^2(\bA^{\text{H}}\bA)^{-1}).
\end{aligned}
\label{dist}
\end{equation}
The Gaussian distribution in \eqref{dist} shows that the estimator is unbiased. Consequently, the MSE of this estimator can be calculated as the trace of the channel estimate covariance matrix, i.e.,
\begin{equation}
\begin{aligned}[b]
\textrm{MSE} = \mathds{E} [(\hat{\bh}-{\bh})^{\text{H}}(\hat{\bh}-{\bh})] = \sigma^2 \text{tr} [(\bA^{\text{H}}\bA)^{-1}].
\end{aligned}
\label{traceA}
\end{equation}

Reference \cite{negi1998pilot} has studied the use of scattered pilots for channel estimation in OFDM systems. The general conclusion drown there is that to minimize the MSE of the channel estimates the pilots should be equally spaced across subcarriers. This conclusion that was derived theoretical in \cite{negi1998pilot} matches ones intuitive understanding as well. Taking note of this we argue that for FBMC also the use of equally spaced pilots should lead to an optimized estimate of the channel. Making this assumption, a detailed study of \eqref{demod} and \eqref{matform} reveals that \eqref{traceA} may be approximated as
\begin{equation}
\textrm{MSE} = \sigma^2 \text{tr} \big[\big( (\bP \BDFT_{N_{\rm p},L} )^{\text{H}}(\bP \BDFT_{N_{\rm p},L}) \big)^{-1}\big],
\label{traceX}
\end{equation}
where $\bP = {\rm diag}(\p)$ and $\BDFT_{N_{\rm p},L}$ is the ${N_{\rm p}}\times L$ matrix holding the first $L$ columns of the DFT matrix of size ${N_{\rm p}}$.  Incidentally, \eqref{traceX} has the same form as the cost function that has been used in \cite{negi1998pilot} to arrive at the conclusion that equally spaced pilot placement is optimum.

Next, we note that rearrangement of \eqref{traceX} leads to
\begin{equation}
\begin{aligned}
\textrm{MSE} &= \sigma^2 {\rm tr} [ (\BDFT_{N_{\rm p},L} ^{\text{H}} \bP^{\text{H}} \bP \BDFT_{N_{\rm p},L})^{-1}] \\
&= \sigma^2  {\rm tr} [ (\BDFT_{N_{\rm p},L} ^{\text{H}} \frac{P_{\rm t}}{N_{\rm p}} \bI_{N_{\rm p}} \BDFT_{N_{\rm p},L})^{-1}] \\
& =\sigma^2  {\rm tr} [ (  \frac{P_{\rm t}}{N_{\rm p}} N_{\rm p} \bI_{L} )^{-1}] =\frac{\sigma^2 L}{P_{\rm t}}.
\label{traceXF}
\end{aligned}
\end{equation}
where $P_{\rm t}$ denotes the total power of the pilots. This shows that the MSE of the channel estimate only relates to the total transmit power of the pilots. Accordingly, as long as the total power of the pilot symbols is constant and the number of pilots is equal or larger than the length of the channel, using a smaller number of pilots than $M$ does not degrade the channel estimation accuracy. Therefore, a pilot structure that minimizes the MSE of the channel estimate has to follow two rules; (i) distribute pilots with the same distance, and (ii) keep the total power fixed.

\label{conventionalmethod}

\vspace{-1.4mm}
\section{Multiuser Channel Estimation}
\label{sec:Multiuser_ch_estimation}

\begin{figure}[t]
    		\psfrag {time}{\hspace{0mm} \small Time }
		\psfrag {freq}{\hspace{0mm} \rotatebox{90}{\small Frequency} }
		\psfrag {red}{\hspace{0mm} \scriptsize User $1$ pilots }
		\psfrag {blue}{\hspace{0mm}   \scriptsize User $2$ pilots }
		\psfrag {green}{\hspace{0mm}   \scriptsize User $U$ pilots }
		\psfrag {zero}{\hspace{1mm}\scriptsize Guard }
		\psfrag {data}{\hspace{1mm}\scriptsize  Data }
		\psfrag {a}{\hspace{0mm} $\vdots$ } 
		\psfrag {b}{\hspace{0mm} $\vdots$ }
		\psfrag {c}{\hspace{0mm}  $\vdots$}
	         \psfrag {d}{\hspace{0mm}  $\vdots$}
		\psfrag {e}{\hspace{0mm} $\vdots$ }	
		\psfrag {f}{\hspace{0mm} $\dots$ }
		\psfrag {g}{\hspace{0mm} $\dots$ }
		\psfrag {h}{\hspace{0mm} $\dots$ }
		\psfrag {i}{\hspace{0mm} $\dots$ }
		\psfrag {j}{\hspace{0mm} $\dots$ }
		\psfrag {k}{\hspace{0mm} $\dots$ }
		\psfrag {l}{\hspace{0mm} $\dots$ }
		\psfrag {m}{\hspace{0mm} $\dots$ }
		\psfrag {n}{\hspace{0mm} $\dots$ }
		\centering	\includegraphics[scale=0.38]{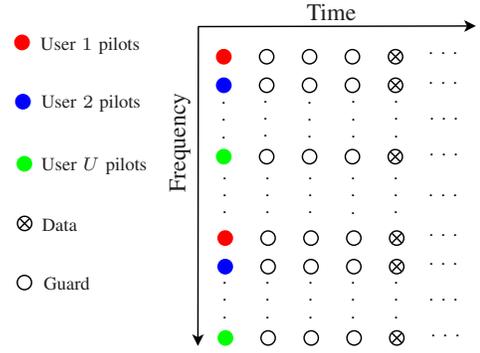}
		\caption{Proposed pilot structure for channel estimation in the uplink.}
		\label{pilot2user}
	\end{figure}

As mentioned earlier, the requirement for guard symbols to guarantee orthogonality of multiple users' pilots in the uplink, makes most of the available solutions in the literature spectrally inefficient. To address this issue, here, we extend the single user time domain channel estimation technique that was discussed in Section~\ref{sparsepilot} to multiuser uplink scenario. We show that through application of our proposed solution, the guard symbols between different users' pilots can be completely removed. Additionally, similar to the single user scenario, in our proposed technique, $L_{u}$ pilots for a given user $u$ are deployed, where $L_{u}$ is the channel length for user $u$. It is worth mentioning that time-frequency resources being utilized for pilot signals of different users are mutually exclusive. In other words, each time-frequency slot can be only used by the pilot of a given user. Fig. \ref{pilot2user}  presents our proposed pilot structure for channel estimation in the uplink. Moreover, as noted earlier, With the proposed pilot structure, the PAPR at the output of each transmitter is reduced compared to the method of \cite{kofidis2015preamble} and \cite{singh2019time} where pilots are transmitted on all subcarriers.

In the absence of guard symbols between the pilots of different users, intrinsic interference makes estimation of different users channels dependent on each other. Taking note of this point, in the following, we take advantage of the intrinsic interference and propose a channel estimation method in which the base station jointly acquires all the users' channel impulse responses. Through simulation results in Section~\ref{simulation}, we show that our proposed channel estimation method not only improves spectral efficiency, but it also achieves almost the same performance as the Cram\'er-Rao lower bound that we have calculated in the end of this section.

 We consider a single base station serving $U$ users in the uplink. The demodulated pilot signal at the base station on subcarrier $m$ and time index $n$ can be expressed as 
 \begin{equation}
    \begin{aligned}
   {z}_{m,n} =&
   \sum_{u=1}^{U} \sum_{l=0}^{L_{u}-1} \sum_{k=-\infty}^{\infty} \sum_{m'=0}^{M-1} \sum_{n' =-\infty}^{\infty} s^{u}_{m',n'} g[k-l-n'\frac{M}{2}] \\
   &\times g[k-n\frac{M}{2}] 
   e^{j2\pi(m'-m)k/M} e^{j\pi (m'+n'-m-n)/2} \\
   &e^{-j2\pi m'l/M}h_{u}[l]+\eta_{m,n}. 
    \end{aligned}
      \label{demodap2}
    \end{equation}
    \vspace{-3mm}

where $h_{u}[l]$ represents the channel impulse response of a given user, $u$, with the length $L_u$. This formula shows the effect of the transmit symbol at a given time-frequency slot on its adjacent slots at the receiver output before the real part operation. Let the vector $\bz_u =  [z_{p_1^u},z_{p_2^u}, ... ,z_{p^u_{N^u_{\rm p}}}]^{\textrm{T}} $ include all the received pilots of the user $u$. Using \eqref{demodap2}, we have,
\begin{equation}
\bar{\bz} = \bar{\bA}\bar{\bh}+ \bar{\boldsymbol{\eta}},
\label{matformap2}
\end{equation}
where $\bar{\bz} = [\bz_1^{\rm T},\ldots,\bz_U^{\rm T}]^{\rm T}$, $\bar{\bh} = [\bh_1^{\rm T},\ldots,\bh_U^{\rm T}]^{\rm T}$, $\bar{\boldsymbol{\eta}} = [\boldsymbol{\eta}_1^{\rm T},\ldots,\boldsymbol{\eta}_U^{\rm T}]^{\rm T}$, $\bh_u=[h_u[0],\ldots,h_u[L_u-1]]^{\rm T}$ is the channel vector of user $u$, $\boldsymbol{\eta}_u$ is the noise contribution to the received pilot sequence of user $u$ and,
\begin{equation}
\bar{\bA} =
 \begin{bmatrix}
    \bA_1 & \boldsymbol{\zeta}_1^2 & \hdots &  \boldsymbol{\zeta}_1^U\\
     \boldsymbol{\zeta}_2^1 & \mathbf{A}_2 & \hdots & \boldsymbol{\zeta}_2^U\\
     \vdots & \vdots & \ddots & \vdots \\
     \boldsymbol{\zeta}_U^1 & \boldsymbol{\zeta}_U^2 & \hdots & \mathbf{A}_U \\
\end{bmatrix}.\nonumber
\end{equation}

Matrices $\bA_u$ and $\boldsymbol{\zeta}_{u_1}^{u_2}$ have the sizes  $N_{\rm p}^u \times L_u$ and  $N_{\rm p}^{u_1} \times L_{u_2}$, respectively. Using \eqref{demodap2} and \eqref{matformap2}, the elements of $\bA_u$ and $\boldsymbol{\zeta}_{u_1}^{u_2}$ matrices can be acquired as
       \begin{equation}
    \begin{aligned}
    A_{u}(m,l)&= \!\!\!
    \sum_{k=-\infty}^{\infty} \sum_{m'=0}^{M-1} \sum_{n'=-\infty}^{\infty} \!\! s_{m',n'}^{u}  g[k\!-\!l\!-\!n'\frac{M}{2}]
   g[k-n\frac{M}{2}]  \\
  & \times e^{j2\pi(m'-m)k/M} e^{j\pi (m'+n'-m-n)/2} 
   e^{-j2\pi m' l/M},
    \end{aligned}
    \label{{elementformulaap2}}
    \end{equation}
    and 
          \begin{equation}
    \begin{aligned}
    \zeta_{u_1}^{u_2}({m,l})& = \!\! \!\! \sum_{k=-\infty}^{\infty} \sum_{m'=0}^{M-1} \sum_{n'=-\infty}^{\infty} \!\! s_{m',n'}^{u_2}  g[k\!-\!l\!-\!n'\frac{M}{2}]
   g[k-n\frac{M}{2}]  \\
   &\times e^{j2\pi(m'-m)k/M} e^{j\pi (m'+n'-m-n)/2} 
   e^{-j2\pi m'l/M}.
    \end{aligned}
    \label{{elementformulazeta}}
    \end{equation}
Similar to \eqref{matform}, the matrices $\bA_u$ for $u = 1,\ldots,U$ define the input-output relationship between the transmit and receive pilot sequences of the users $u$ in the absence of intrinsic interference from other users. The matrices $\boldsymbol{\zeta}_{u_1}^{u_2}$ indicate the intrinsic interference produced by the pilot sequences of the user $u_2$ on the received pilot sequences of the the user $u_1$. Similar to \eqref{matform}, linear equation in \eqref{matformap2} decouples the users' channel responses from their training sequences that are filtered by the transmit and receive filters. This equation considers the intrinsic interference that is caused by non-orthogonal pilots that are deployed by the users. This enables accurate estimation of all the users' channel impulse responses. Hence, here, we take advantage of intrinsic interference between the users' pilot sequences in development of our proposed multiuser channel estimation technique. In presence of guard symbols that completely isolate different users' pilot symbols, the matrices $\boldsymbol{\zeta}_{u_1}^{u_2}$ are equal to zero matrices. As a result, multiuser channel estimation in the uplink reduces to isolated single user channel estimation problems. This comes in expense of a large amount of bandwidth efficiency loss.

Non-orthogonality of adjacent signals in FBMC leads to noise correlation. Therefore, the assumption of uncorrelated noise is not valid in such systems. Presence of guard symbols in sparse pilot structures removes imaginary interference. Consequently, noise correlation between samples used for estimation can be ignored. In contrast, absence of guard symbols in our proposed estimation method turns the covariance matrix of noise into a non-diagonal one. This necessitates consideration of noise correlation in our channel estimation technique. Therefore, we need to modify least squares estimator to include the correlation matrix of channel noise. As the first step, correlation between noise components can be calculated as
    \begin{equation}
    \begin{aligned}
    &{\rm Cov}[\eta_{m,n},\eta_{m',n'}] = \mathds{E} [\eta_{m,n}\eta_{m',n'}^*] -  \mathds{E} [\eta_{m,n}] \mathds{E} [\eta_{m',n'}^*]\\
    &   =  \sigma^2 \sum_{k=-\infty}^{\infty}   g_{m,n}[k] g^*_{m',n'}[k]e^{j2\pi(m'-m)k/M}
        e^{j\pi (m'-m)/2} \\
       &= \sigma^2  \zeta_{m,n}^{m',n'}. 
    \end{aligned}
    \end{equation}
    Hence, the $N_{\rm p}^{\rm t} \times N_{\rm p}^{\rm t}$ noise correlation matrix can be represented as
    \vspace{-1mm}
        \begin{equation}
        \begin{aligned}
 &\bC=  \begin{bmatrix}
    \sigma^2 \zeta_{m_1,n_1}^{m_1,n_1} & \sigma^2 \zeta_{m_1,n_1}^{m_2,n_2} &\dots  & \sigma^2 \zeta_{m_1,n_1}^{m_{N_{\rm p}^{\rm t}},n_{N_{\rm p}^{\rm t}}}\\
     \sigma^2 \zeta_{m_2,n_2}^{m_1,n_1} & \sigma^2 \zeta_{m_2,n_2}^{m_2,n_2} &\dots  & \sigma^2 \zeta_{m_2,n_2}^{m_{N_{\rm p}^{\rm t}},n_{N_{\rm p}^{\rm t}}} \\
    \vdots & \vdots  & \ddots & \vdots \\
     \sigma^2 \zeta_{m_{N_{\rm p}^{\rm t}},n_{N_{\rm p}^{\rm t}}}^{m_1,n_1} & \sigma^2 \zeta_{m_{N_{\rm p}^{\rm t}},n_{N_{\rm p}^{\rm t}}}^{m_2,n_2} &\dots  & \sigma^2 \zeta_{m_{N_{\rm p}^{\rm t}},n_{N_{\rm p}^{\rm t}}}^{m_{N_{\rm p}^{\rm t}},n_{N_{\rm p}^{\rm t}}}
\end{bmatrix} 
\end{aligned}
    \end{equation}
where ${N_{\rm p}^{\rm t}} = \sum_{u=1}^{U} {N_{\rm p}^u}$ is the total number of pilot symbols for all the users. Here, $(m_i,n_i)$ indicates the time-frequency indices of the $i^{\rm th}$ element in $\bar{\bz}$. Thus, the order of the samples in $\bar{\bz}$ determines the structure of the covariance matrix.     

Using  equation \eqref{matformap2}, and considering noise correlation, the least squares channel estimates can be obtained as
       \begin{equation}
       \hat{\bh}_{\rm t} = (\bar{\bA}^{\text{H}} \bC^{-1} \bar{\bA})^{-1} \bar{\bA}^{\text{H}} \bC^{-1} \bar{\bz},
       \label{estap2}
    \end{equation} \vspace{-1mm}
where the vector $\hat{\bh}_{\rm t} = [\hat{\bh}_1^{\rm T},\ldots, \hat{\bh}_U^{\rm T}]^{\rm T}$ includes all the users' channel estimates and $\hat{\bh}_u$ is the channel estimate for a given user $u$. From \eqref{estap2}, one may realize that different users' pilot symbols contribute to intrinsic interference and consequently to the channel estimation of a given user $u$. It is worth noting that the matrix in \eqref{estap2} does not need to be calculated each time. This matrix can be pre-calculated and saved in the memory for an efficient implementation.

In the rest of this section, we calculate CRLB for our proposed technique. We analytically show that the MSE of our proposed estimation method matches the CRLB. Considering $\hat{\bh}_{\rm t}$ as an unbiased estimate, CRLB is defined as \cite{kay1993fundamentals}
\begin{equation}
{\rm Var}[\hat{\bh}_{\rm t}-\bar{\bh}] \geq  {\rm CRLB}=\text{tr}(\bF^{-1}),
\end{equation}
where $\bF$ is the Fisher information matrix. This matrix can be calculated as \cite{kay1993fundamentals},
\begin{equation}
\bF = -\mathds{E} \big[ \frac{\partial^2 \text{ln}({\rm Pr}(\bar{\bz};\bar{\bh})) }{\partial \bar{\bh} \partial \bar{\bh}^{\rm H}}     \big].
\label{fisher}
\end{equation}
Using \eqref{matformap2} in \eqref{fisher}, Fisher information matrix for our proposed estimation scheme is
\begin{equation}
\bF =  (\bar{\bA}^{\rm H} \bC^{-1} \bar{\bA}).
\end{equation}
By definition, CRLB can be calculated as
\begin{equation}
 {\rm CRLB}={\rm tr}(\bF^{-1}) = {\rm tr}[( \bar{\bA}^{\rm H} \bC^{-1} \bar{\bA})^{-1}].
 \label{eq:CRLB}
\end{equation}

For the proposed estimator to reach CRLB, MSE needs to be equal to CRLB. Using \eqref{estap2} and following a similar line of derivations as in Section~\ref{sparsepilot}, the MSE performance of our proposed estimation method in this section can be obtained as 
\begin{equation}
\begin{aligned}[b]
\textrm{MSE} = \mathds{E} [(\hat{\bh}_{\rm t}-\bar{\bh})^{\text{H}}(\hat{\bh}_{\rm t}-\bar{\bh})] =  \text{tr} [(\bar{\bA}^{\text{H}} \bC^{-1}\bar{\bA})^{-1}].
\end{aligned}
\label{eq:MSE_multiuser}
\end{equation}
Based on \eqref{eq:MSE_multiuser} and \eqref{eq:CRLB}, analytical MSE of our proposed technique matches the CRLB. This result is also confirmed through numerical simulations in Section~\ref{simulation}.

\section{Extension to Massive MIMO}\label{sec:Massive_MIMO}
The pivotal principle behind massive MIMO systems relies on the fact that the signals of different users can be distinguished from one another through the channel responses between each user antenna and base station antennas. Considering a large number of antennas being deployed at the base station, multiuser detection can be performed through simple matched filter receivers \cite{rusek2012scaling}. This necessitates accurate knowledge of all the channel responses at the base station. Therefore, channel estimation is of a great importance in massive MIMO systems. FBMC has been recently proposed for application to massive MIMO channels \cite{aminjavaheri2018filter}, \cite{farhang2014filter}. However, to the best of our knowledge, there is no work in the literature discussing channel estimation aspects of FBMC in such environments.

As it was mentioned earlier, extension of the available channel estimation techniques in the literature to multiuser channels leads to a substantial amount of bandwidth efficiency loss if at all they are applicable to massive MIMO. This is due to the requirement of a large number of guard symbols to separate different users' pilots. Additionally, short coherence time of the channel in massive MIMO systems does not allow deployment of orthogonal pilot sequences for all the users in a multicell scenario. This leads to the so called pilot contamination problem in such networks, \cite{jose2009pilot}. A natural solution to combat this issue is through development of efficient channel estimation methods with very small training overheads. 

Assuming synchronous uplink transmission, the channel estimation technique that was developed in Section~\ref{sec:Multiuser_ch_estimation} is a perfect solution to the above problem. In our solution, there is no requirement for insertion of guard symbols between the pilots of different users. Additionally, each user utilizes only $L_u$ pilots. Hence, training pilots are inserted in a very compact structure enabling joint estimation of all the users' channel responses at each base station antenna. Due to the training overhead reduction that is  achieved using our proposed solution, a larger number of orthogonal pilots can be assigned to the users in adjacent cells. This will relax the pilot contamination problem in FBMC-based massive MIMO systems. To highlight the benefits of our proposed channel estimation method in massive MIMO channels, in the next section, we evaluate its sum-rate performance in comparison with the conventional time domain channel estimation technique in \cite{kong2014time}.
\vspace{-1.5mm}
 \section{Simulation Results}
 \vspace{-1mm}
    \label{simulation}
In this section, we evaluate the normalized MSE  (NMSE) and rate performance of our proposed channel estimation method through computer simulations. For all simulations, we consider $M=128$  and 4-QAM (quadrature amplitude modulation) signaling. We use PHYDYAS  prototype filter, \cite{bellanger2010fbmc}, with overlapping factor $\kappa=4$.  A random channel model with exponential power delay profile of $\alpha_k(l) = e^{-\beta_k l }$  with independent Raleigh fading paths and $L=32$ is considered. To make the results comparable in different multiuser scenarios, for the user of interest, $\beta_k=0.5$ and for other users, $\beta_{k'} \in [0.4, 0.6]$ are considered. For isolation of preamble from data symbols and to avoid intrinsic interference issues, three guard symbols , i.e., $\kappa-1$, are inserted between them in time. To obtain
the NMSE results, we have used 1000 independent realizations of the channel. \vspace{-2.2mm}
\subsection{Estimation Performance}
\begin{figure}[t]
		\centering
		\includegraphics[scale=0.59 ,trim={0 0 0 0},clip]{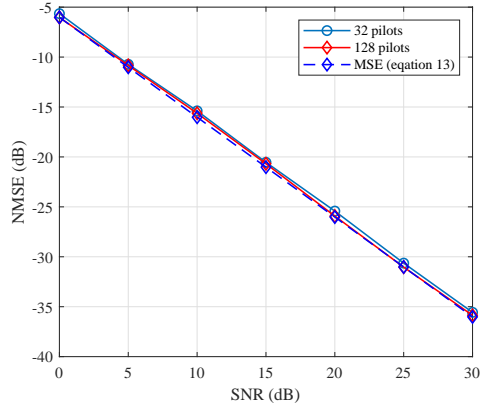}
		\caption{NMSE performance of the proposed technique for single user scenario.}
		\label{128vs32power}
	\end{figure}

In Fig.~\ref{128vs32power}, we study the effect of the number of pilots on estimation performance in terms of NMSE, defined as
\begin{equation}
    {\rm NMSE} = \frac{\sum\limits_{l=0}^{L-1}|\hat{h[l]}-h[l]|^2}{\sum\limits_{l=0}^{L-1}|h[l]|^2}.
\end{equation}
We also confirm the validity of \eqref{traceXF} by numerical results while using it as a benchmark in our performance analysis. To illustrate the effect of our proposed pilot structure with the minimum number of pilots, we first consider the single user scenario for different number of pilots for a fixed training power budget. In Fig. \ref{128vs32power}, the results are presented for the two choices of $N_{\rm P} = M$ and $M/4$. As one would expect from the theoretical results in Section \ref{sparsepilot}, the number of pilots has a negligible effect on channel estimation accuracy. Moreover, the MSE of channel estimates matches the results from equation \eqref{traceXF}.

\begin{figure}[t]
		\centering
		\includegraphics[scale=0.6,trim={0 0 0 0},clip]{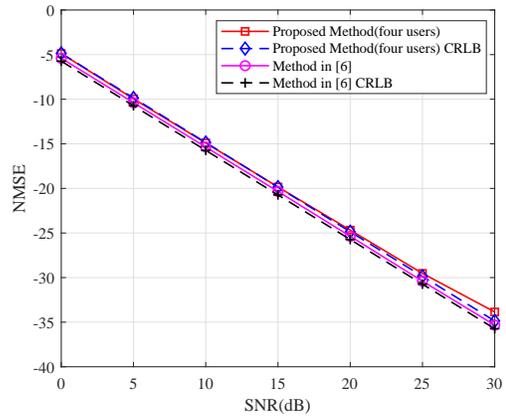}
		\caption{NMSE performance comparison of the proposed technique compared with the one in \cite{kong2014time} for multiuser scenario with $U=4$ users.}
		\label{NMSE}
	\end{figure}

Fig. \ref{NMSE} illustrates the performance of our proposed channel estimation method where four users are present in a cell. These results are for the case where $N_{\rm p}=L=32$. This figure shows that our proposed channel estimation technique reaches the CRLB performance that was derived in Section~\ref{sec:Multiuser_ch_estimation}. The results are also compared with the time domain approach in \cite{kong2014time} and found to be about the same. 
\vspace{-2mm} \subsection{Sum-Rate Performance}
To confirm our claims in Section~\ref{sec:Massive_MIMO}, here, we adopt ideal uplink sum-rate calculations considering training overheads to better illustrate the benefits of our proposed method in massive MIMO. We first consider a single cell setup with $U=4$ users and $128$ receive antennas at the base station where maximum ratio combining (MRC) is deployed for detection. To calculate the sum-rate, we use the following formula.\vspace{-1.5mm}
\begin{equation}
    {\rm \text{sum-rate}} = \gamma \sum_{u=1}^{U} {\rm log}_2 (1+{\rm SINR}_u),
\end{equation}
where ${\rm SINR}_u$ represents the signal to interference plus noise ratio (SINR) at the receiver output for the $u$th user and $\gamma$ is the rate loss due to the training overhead. 

\begin{figure}[t]
		\centering
		\includegraphics[scale=0.56,trim={0 0 0 0},clip]{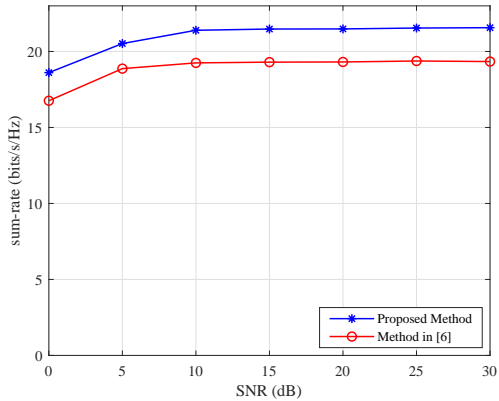}
		\caption{Per cell sum-rate performance comparison of our proposed method with the method of \cite{kong2014time} in a single cell scenario.}
		\label{to_sumrate}
	\end{figure}

Fig. \ref{to_sumrate} depicts the sum-rate as a function of SNR. Here, we have set $N_{\rm p}=L=16$, $M=128$, have assumed a channel coherence interval equivalent to the duration of 84 FBMC symbols, and accordingly chosen a packet length of 84 FBMC symbols. Each packet starts with pilots and guard symbols as a preamble and the rest of the packet is filled with data symbols. 

Fig. 4, further, compares sum-rate performance of our proposed solution with \cite{kong2014time} which uses $N_{\rm p}=128$ to estimate each user's channel, i.e., it has four times the training overhead compared to our proposed method. 

\begin{figure}[t]
		\centering
		\includegraphics[scale=0.56,trim={0 0 0 0},clip]{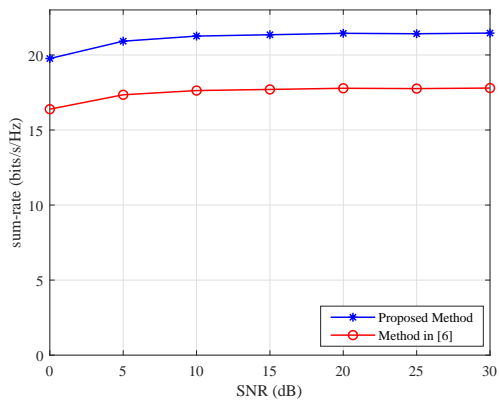}
		\caption{Per cell sum-rate performance comparison of our proposed method with the method of \cite{kong2014time} in a two cell scenario.}
		\label{2cellsumrate}
	\end{figure}

To illustrate the effect of the proposed method in a multicell scenario, similar to [4], we consider a setup with two cells. Each cell serves four users with  $N_{\rm p}=L=16$ and a random cross-gain $g \in [0, 1]$. The cross-gain factors between each user in neighbor cell and the base station antennas of the cell of interest may be thought as path loss coefficients. Our proposed method is capable of serving a total of $8$ users in two cells with the channel length $L=16$. The sum-rate is depicted in Fig. \ref{2cellsumrate}. Similar to Fig. \ref{to_sumrate}, a substantial gain due to training overhead reduction is evident.

\section{Conclusion}
\label{sec:Conclusion}
In this paper, we proposed a novel time domain multiuser channel estimation method for the uplink of FBMC-based systems. We analytically proved the optimality of our proposed estimator by showing that its MSE expression is the same as the derived CRLB. The proposed method requires a small number of pilots, equal to the respective channel length, for each user. This leads to a significant amount of savings in training overhead. We proposed a pilot structure where pilots for different users are interleaved in time and frequency without the need for any guard symbols between them. This brings additional savings in training overhead which becomes substantial as the number of users increases. Accordingly, the proposed method significantly improves spectral efficiency compared to the available solutions in the literature. Compared to the methods
that transmit pilots on all the subcarriers, sparsity of our proposed pilot
structure leads to a reduced PAPR. We also discussed extension of our solution to massive MIMO and showed that its reduced training overhead relaxes the pilot contamination issue in multicell massive MIMO channels. Finally, we confirmed the efficacy of our proposed method and analytical derivations through simulations. 
\section{Acknowledgement}
The portion of this research that has been performed at the University of Utah is supported through the National Science Foundation grant SpecEES-1824558. 

\ifCLASSOPTIONcaptionsoff
  \newpage
\fi

%


\end{document}